\documentstyle{mn} \title[K-band polarimetry of 
three high-redshift
radio galaxies]{Upper limits on $K$-band polarization in three
high-redshift radio galaxies: 53W091, 3C 441 and MRC 
0156$-$252} \author[G. Leyshon, J. S.
Dunlop and Stephen A. Eales]{G.  Leyshon,$^{1,3}$ J. S. Dunlop$^{2,4}$ and
Stephen A. Eales$^{1,5}$\\ $^1$Department of Physics and Astronomy,
      Cardiff University,
      P.O. Box 913,
      Cardiff CF2 3YB\\
$^2$Institute for Astronomy, Department of Physics and Astronomy,
      The University of Edinburgh, 	
      Edinburgh EH9 3HJ\\
$^3${\verb"G.Leyshon@astro.cf.ac.uk"}\hspace{0.8cm}
$^4${\verb"J.Dunlop@roe.ac.uk"}\hspace{0.8cm}
$^5${\verb"S.Eales@astro.cf.ac.uk"}
}

\input{psfig.sty}

\DeclareSymbolFont{UPM}{U}{eur}{m}{n}
\SetSymbolFont{UPM}{bold}{U}{eur}{b}{n}
\DeclareMathSymbol{\umu}{0}{UPM}{"16}
 
\newcommand{\micron}{$\umu$m}

\newcommand{\cduo}[1]{\multicolumn{2}{c}{#1}}

\newcommand{\tbn}[1]{${^{#1}}$}

\newcommand{\Pq}[1]{{P}_{{\rm{Q,}}#1}}
 
\newcommand{\ccen}[1]{\multicolumn{1}{c}{#1}}

\newcommand{\Ptt}[1]{{P}_{#1}} 
 
\newcommand{\PHI}[1]{{\Phi}_{#1}}

\newcommand{\Kb}{{\it K}}

\begin{document}

\maketitle

\begin{abstract}

 We present the results of \Kb-band imaging polarimetry of three radio
galaxies, including the very red and apparently old $z = 1.55$ 
galaxy 53W091. We find weak evidence for
polarization in components of 3C 441 and in the south-east companion of
53W091, but no evidence of significant polarization in 53W091 itself.
We also find strong evidence that MRC 0156$-$252 is unpolarised. We present
upper limits for the \Kb-band polarization of all three sources. For
53W091, the lack of significant \Kb-band polarization provides further
confidence that its red $R-K$ colour can be attributed to a mature
stellar population, consistent with the detailed analyses of its
ultraviolet spectral-energy distribution which indicate a minimum age of
2$-$3.5 Gyr.

\end{abstract}


\section{Introduction}

 The visible and infrared light emitted by radio galaxies is thought to be
a combination of starlight and nebular emission from the host galaxy, and
a scattered power-law component originating in the active nucleus hidden
in the heart of the galaxy.  Manzini \& di Serego Alighieri
\shortcite{Manzini+96a} have examined a small sample of radio galaxies at
redshifts ranging from 0.11 to 2.63, and have demonstrated that their
observed magnitudes (by multiwaveband photometry) are consistent with
spectra synthesised from three such components. 

 The Unified Model of radio-loud quasars and radio galaxies \cite[and references
therein]{Robson-96a} suggests that both species contain an active nucleus
consisting of an accretion disc shrouded in a dusty torus, with jets
emerging at either pole. Those sources with jets oriented perpendicular to
our line of sight are seen as radio galaxies with extended radio lobes but
with their optical nuclear light shrouded, while those viewed `down the jet' are
classed as brilliant quasars with strong emission lines. It is now thought
that some apparent `radio galaxies', such as 3C 22
\cite{Rawlings+95a,Dunlop+93a},
may be in the quasar orientation, but with opaque material obscuring
much of the light from the active nucleus. 

It has been known for a decade that visible light from high-$z$ radio
galaxies is often aligned with the radio axis
\cite{Chambers+87a,McCarthy+87a} -- the so-called `alignment effect' --
and that this light is often polarized with its $\bmath{E}$-vector
oriented perpendicular to the radio axis
\cite{Alighieri+89a,Jannuzi+91a,Tadhunter+92a}. Both of these effects
can be explained if the visible light is dominated by the nuclear
component:  nuclear light escapes from the active nucleus in a narrow
cone about the radio jet, and a fraction is scattered towards us by
dust or electrons. The scattering region forms an extended optical
structure approximately parallel to the radio jet, and the scattered
light becomes polarised perpendicular to the optical structure in the
scattering process.  Recent Keck spectropolarimetry by Cimatti et al.\
\shortcite{Cimatti+96a,Cimatti+97a,Cimatti+98a}, Dey et al.\
\shortcite{Dey+96a} and Tran et al.\ \shortcite{Tran+98a} have
confirmed the presence of continuum light and broad lines in the
polarized spectra of radio galaxies at $1.0 \la z \la 2.5$, with the
polarization orientation being approximately perpendicular to the
optical structure.

Manzini \& di Serego Alighieri \shortcite{Manzini+96a} model radio galaxy
spectra at rest frame wavelengths from 0.2 \micron\ to 1.0 \micron. The
contribution of the starlight becomes greater at longer wavelegths, while
the scattered nuclear component decreases. For five out of their six
galaxies, the stellar component of the light has become dominant by a
rest-frame wavelength of 0.5 \micron; in 3C 277.2 $(z=0.766)$ the
starlight only exceeds the nuclear component at about 0.85 \micron.

 \Kb-band observations of radio galaxies at $z \sim 1$ image light 
emitted at $\lambda \sim 1.1$ \micron\ in the rest frame. It is not 
obvious, {\em a priori}, that non-stellar features will be visible in 
this band; in practice, however, the alignment effect has been detected in 
the infrared \cite{Dunlop+93a}, albeit
sometimes weakly \cite{Rigler+92a}. It is now known that a substantial 
fraction of the $K$-band emission arises in the active nuclei of the most 
radio-loud galaxies, namely 3C sources and their southern hemisphere 
equivalents \cite{Eales+97a}.

 The first measurements of the \Kb-band polarizations of high-redshift 
radio galaxies were carried out by Leyshon and Eales \shortcite{Leyshon+98a} 
(hereafter L\&E) in 1995. They found evidence for polarization in 3 out 
of 7 galaxies, and tentative evidence in two further cases. 
The galaxies formed a representative sub-sample of the 3C catalogue: 
polarization values ranged from 3 to 12 per cent, suggesting that up to 
a third of the \Kb-band emission from such galaxies could originate in the 
active nucleus.

 In their first study, L\&E found that their two optically
compact sources (3C 22 and 3C 41) gave the expected perpendicular
alignment, but their extended sources (3C 114 and 3C 356) gave roughly
parallel alignments, within rather large directional error bars. In this 
paper we continue our programme of \Kb-band polarimetry with further 
observations of 3C 441, and polarimetry of MRC 0156$-$252 and the 
very red ($R-K = 5.8$) $z = 1.55$ milli-Jy radio galaxy LBDS 53W091.

\section{Target Objects}

The study of \Kb-band polarizations of distant, faint, radio galaxies
is very much in its infancy. Our first paper (L\&E) concentrated on 3C radio 
galaxies at $z \sim 1$. In this new study, we have extended our 
investigation to higher redshifts, choosing two targets with very 
different radio luminosities.

 Galaxy LBDS 53W091 has a very low radio luminosity, which suggests the 
presence of a very weak active nucleus; but there is an unpublished claim 
(Chambers, private communication) of a $\sim 40$ per cent 
$H$-band polarization. Radio galaxy MRC 0156$-$252 has a high radio 
luminosity and infrared spectroscopy (see below) suggests that a large 
fraction of its $K$-band light may originate in an active nucleus. We 
also observed 3C 441, for which previous observations (L\&E) had returned 
marginal results.

\subsection{LBDS 53W091}

\begin{figure*}
\begin{minipage}{165mm}
\psfig{file=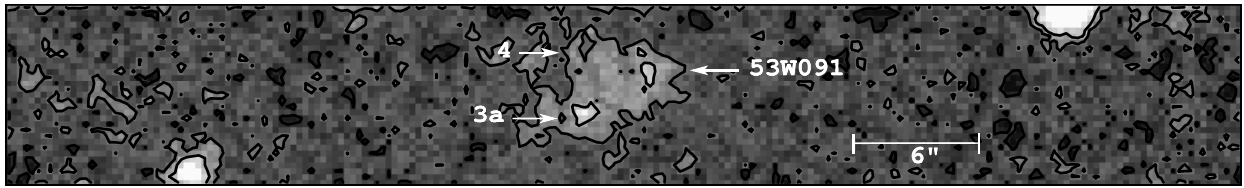,width=165mm}
\caption{53W091 - stacked \Kb-band image of August 1997 and July 1997 data, 
total integration time 5 hours 56 minutes. [4 hours 4 minutes (August) plus 
1 hour 52 minutes (July).] Overlaid with contours at 8\%, 16\% and 24\% of the 
peak intensity present in the image. 
  (North is at the top, East at the left.)}
\label{Wpic} \end{minipage} \end{figure*}

 The galaxy LBDS 53W091 (Figure \ref{Wpic}) has aroused great 
excitement in the last two
years. First investigated by Dunlop et al.\ \shortcite{Dunlop+96a} as an
extremely red radio-source identification, it was found to be a 
very red radio galaxy
visible at high ($z=1.552$) redshift. 
Its spectrum exhibits
late-type absorption features, and no prominent
emission features. 
The lack of emission features suggests that the active nucleus responsible
for its $\sim 25$ mJy 1.4 GHz radio emission contributes very little light
to the visible/ultraviolet, and consistent with this 
Spinrad et al. \shortcite{Spinrad+97a} note that radio galaxies with 
weak active nuclei ($S_{\rm 1.4GHz}<$ 50 mJy) generally are not expected to 
be dominated by optical nuclear emission, and do not display the 
alignment effect (their 4.86 GHz radio map of 53W091 reveals a 
double-lobed FR-II steep spectrum radio source, where the radio lobes are 
separated by approximately 4\farcs3 at position angle 131\degr).

Dunlop et al.\ \shortcite{Dunlop+96a} argue
that the galaxy is unlikely to be an obscured quasar and that its red
colour $R-K = 5.8$ is due to its mature stellar population.
This is based on a detailed investigation of its rest-frame ultraviolet
spectral energy distribution. This appears essentially identical to that of an
F5V star, indicating that stars of spectral type younger
than approximately F2 have all evolved off the main-sequence
\cite{Dunlop+96a}.

The importance of 53W091 is therefore that, despite being initially
discovered as a weak radio source, it is currently the best example
of a red elliptical at $z > 1.5$ whose colours appear to be
entirely due to an `old' stellar population. Thus, irrespective of its
precise age, this galaxy provides an
apparently robust datapoint on the red envelope of galaxy evolution.
The quality of the Keck spectrum of 53W091 has encouraged several authors
to attempt a detailed age determination of its stellar population. The
initial analysis by Dunlop et al.\ \shortcite{Dunlop+96a} indicated an 
age in the range
$3-4$ Gyr (for solar metallicity), and this was supported by the more detailed 
investigation of Spinrad et al.\ \shortcite{Spinrad+97a}. Such an age is 
only consistent with its high
redshift in certain cosmologies, requiring a low-density Universe 
$(\Omega_0 \sim 0.2)$, or 
else an unacceptably low Hubble constant for critical density.
Subsequently, Bruzual \& Magris \shortcite{Bruzual+98a}
concluded in favour of a younger age based on the most recent
models of Bruzual \& Charlot, although detailed fitting with these models
still indicates an age $> 2.5$ Gyr \cite{Dunlop-99a}. More recently Yi 
\shortcite{Yi-99a} has concluded in favour of a minimum age of 2 Gyr. 
However, this
age is derived from a model which has a metallicity of twice solar, and
if solar metallicity is assumed the age rises to $> 3$ Gyr. In summary,
detailed age dating currently indicates that between 2 and 3.5 Gyr have
elapsed since the last major epoch of star formation in 53W091, but
irrespective of its precise age, 53W091 appears to be the best available
example of a passively evolving elliptical galaxy at $z \simeq 1.5$.

Given this background, it therefore came as a major surprise
when unpublished polarimetric data was recently reported as indicating
that 53W091 has a very high infrared polarization -- 
of order 40 per cent (Chambers, private
communication). This measurement suggested that very little of its observed
infrared light was contributed by stars, a result completely at odds with
the predicted infrared luminosity of the stellar population based on the
above-mentioned analyses of the ultraviolet SED.
A major objective of the observations presented here was therefore to
attempt to confirm or refute this result.

\subsection{MRC 0156$-$252}

 Eales \& Rawlings \shortcite{Eales+96a} have compared radio galaxies at
redshifts $z \sim 1$ and $z > 2$, and find that those radio galaxies at
$z>2$ have brighter absolute $V$-band magnitudes. Such a result might be
attributed either to strong evolution in the stellar population of radio
galaxies, or to an increasing misclassification of quasars as radio
galaxies at higher redshifts. 

 MRC 0156$-$252 has the brightest known absolute $V$-band magnitude for a
radio galaxy at $z \sim 2$. The cause of its high luminosity is uncertain:
it may be being viewed during an epoch of intense star formation, or 
it may be a quasar obscured by dust \cite{Eales+96a}.
 Broad H$\alpha$ lines suggest that some of its
light is originating in an active nucleus. McCarthy et al.
\shortcite{McCarthy+90a} earlier classified it as a radio galaxy and
suggested \cite{McCarthy-93a}, on the basis of its red spectral energy
distribution, that it was a galaxy at an advanced stage of evolution.
The radio flux is 131 $\pm$ 13 mJy at 4.85 GHz \cite{Griffith+94a} and
1.39 $\pm$ 0.05 Jy at 408 MHz \cite{Large+81a}.

\begin{figure*}
\begin{minipage}{165mm}
\psfig{file=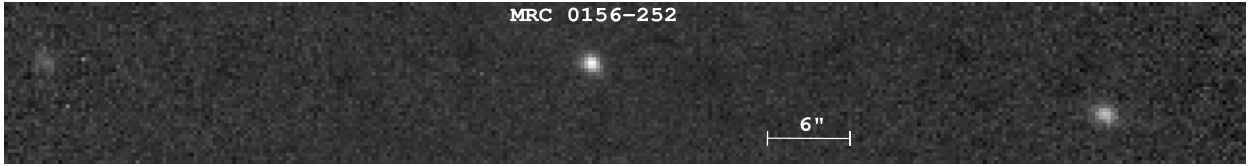,width=165mm}
\caption{MRC 0156$-$252 - stacked \Kb-band image of August 1997 data, 
total integration time 3 hours 44 minutes - including one cycle of 
observations which was not used for our subsequent polarimetry.  (North 
is at 
the top, East at the left.)} \label{Mpic} \end{minipage} \end{figure*}

MRC 0156$-$252 appears unresolved in our \Kb-band image (Figure \ref{Mpic}),
verifying the findings of McCarthy, Persson \& West
\shortcite{McCarthy+92a}, who did, however, find extended structure in
their visible-band images. Our polarimetry allows us to investigate
whether \Kb-band light has been scattered into our line of sight. 

\subsection{3C 441}

\begin{figure*}
\begin{minipage}{165mm}
\psfig{file=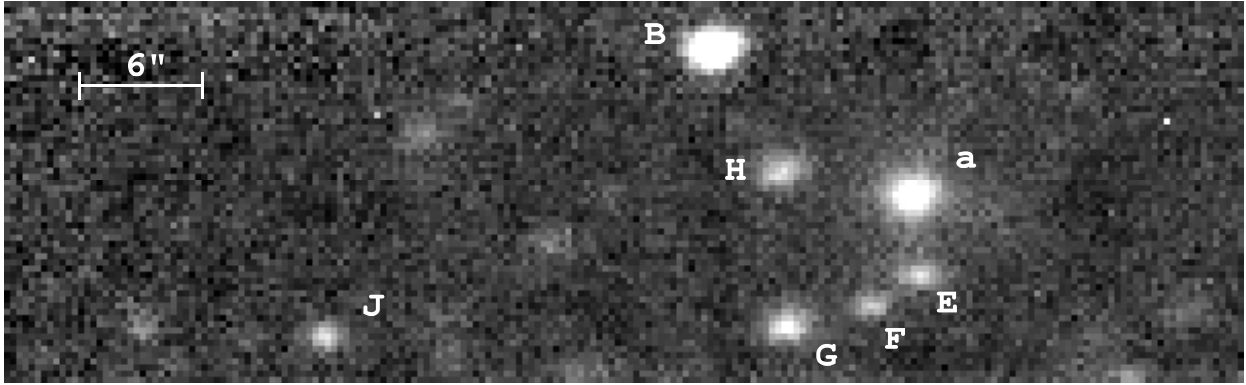,width=165mm}
\caption{3C 441 - stacked \Kb-band image of August 1995 and August 1997 data, 
total integration time 4 hours 16 minutes. [1 hour 52 minutes (1997) plus 2 
hours 24 minutes (1995).]  (North is at the
top, East at the left.) The objects are labelled as detailed in the text.}
\label{Cpic} \end{minipage} 
\end{figure*}

3C 441 appears in a rich field (Figure \ref{Cpic}) with five neighbours; 
identification of the radio core as object $a$ is based on the
observations of Riley, Longair \& Gunn \shortcite{Riley+80a} and is
apparently confirmed by the work of McCarthy \shortcite{McCarthy-88a} and
of Eisenhardt \& Chokshi \shortcite{Eisenhardt+90a}. The object labelled
$a$ follows the notation of Lacy et al. \shortcite{Lacy+98a}, who identify
objects $c$ and $d$ to the north-west of object $a$ -- these are not shown
in the restricted declination of Figure \ref{Cpic}, though object $c$ was
clearly visible in the data taken by L\&E.  The star $B$ is labelled as in
Riley et al. \shortcite{Riley+80a}, and the remaining objects are labelled
$E$ thru $J$. 

 Fabry-Perot imaging by Neeser
\shortcite{Neeser-96a} shows that none of the other objects in the field
lie within a velocity range $(-1460, +1180)$ km\,s$^{-1}$ of 3C 441 itself,
but Neeser \shortcite{Neeser-96a} questions whether the identification of
3C 441 is correct - arguing that it may in fact be our object $F$. 

The recent 8 GHz radio map of 3C 441 \cite{Lacy+98a} supports the 
traditional identification of the radio core with object 
$a$, and shows that the North-West jet is deflected to 
the south where it would otherwise have encompassed object $c$.
It is known that 3C 441 has a broad-band optical polarization which is
roughly perpendicular to its radio structure \cite{Tadhunter+92a}.

\section{Instrumentation and reduction procedure}

 Polarimetric observations were obtained using
IRPOL2, installed at UKIRT (the United Kingdom Infrared
Telescope, Hawaii) as described in L\&E. We took data on the nights of
1997 August 18 and 19. IRPOL2 makes use of a Wollaston prism to split the
light into parallel beams of orthogonal polarization. For our 1997 run
(unlike that documented in L\&E) we made use of the instrument's focal plane 
mask. We created mosaics of each source by combining seven 60-s exposures 
at horizontal spacings of $9 \arcsec$ and vertical spacings of $\pm 1 
\arcsec$. The total exposure time (summed over all four waveplate 
settings) is listed in Table \ref{threeobs}. 

\begin{table*}
\begin{minipage}{165mm}
\caption{Redshifts, integration times and extinctions for our sample of radio
galaxies.}
\label{threeobs}
\begin{tabular}{llrrrrrrr}
\hline 
Source & IAU form & \ccen{$z$} & $\lambda_r$ & $t_{\rm int}$ &
\ccen{$l$} & \ccen{$b$} & $A_{\rm B}$ & $p_K$ \\
LBDS 53W091  & 1721$+$501 & 1.552 & 0.86 & 364 &  76.9 & +34.5 & 0.08 & 0.06\\
MRC 0156$-$252 & 0156$-$252 & 2.016 & 0.73 & 196 & 208.6 & -74.8 & 0.00 & 0.00\\
3C 441       & 2203$+$292 & 0.707 & 1.29 & $\dagger$112 &  84.9 & -20.9 & 
0.34 & 0.24\\ \hline
\end{tabular}

\medskip
 
Key: $z$: redshift; $\lambda_r$ (\micron): rest-frame equivalent of
observed-frame 2.2\micron;
$t_{\rm int}$ (min) : total integration time (min) summed over all waveplate 
settings ($\dagger$: 3C 441 data was pooled with L\&E's 144 minutes from 
August 1995 giving 256 minutes in total) ; $l$ (\degr): 
Galactic longitude ({\sc ned}); $b$ (\degr): Galactic latitude ({\sc 
ned}); $A_{\rm B}$: Blue-band extinction (magnitudes), from {\sc ned}, 
derived from Burstein \& Heiles \shortcite{Burstein+82a}; $p_K$: maximum 
interstellar contribution to \Kb-band polarization (per cent).
\end{minipage}
\end{table*}
 
 Data reduction was performed using the standard UKIRT {\sc ircamdr}
software to flat field the images, and {\sc iraf} to align and combine the
mosaics. The images were then averaged to form a single image for each
waveplate, as this gives the best signal-to-noise in the final polarimetry
\cite{Leyshon-98a,Maronna+92a,Clarke+83a}. The 3C 441 data was then
combined, at each waveplate position, with our stacked 1995 images.
Results are summarised in Table \ref{restab}.

\begin{table*}
\begin{minipage}{165mm}
\caption{Observational results from polarimetry of 3 radio galaxies and 
neighbouring objects in the $K$-band.}
\begin{tabular}{lrrrrrrrrrrrrr}
\hline
Source & $r$(\arcsec) & $q (\sigma_q) $(\%) & $u (\sigma_u)$(\%)
& $P \pm \sigma_P$(\%) & $2\sigma$.UL & Prob &
$\theta$(\degr) & \cduo{$\sigma_\theta$(\degr)} & $\PHI{K}$(\%)\\
\hline 
LBDS 53W091 $\natural$  & 1.1 & 0.08 (16.6) & -7.5 (22.1) & 0.37 
 $\pm$ \tbn{L}7.80 &  30.8 & 10.8 & 38.3 & \cduo{$\pm$59.6} & 25 \\
LBDS 53W091 $\flat$ & 1.1 & 0.64 (16.2) &  -3.42 (18.4) & 0.17 
 \tbn{L,U} & 21.5 & 2.02 & 43.3 & \cduo{$\pm$60.0} & (65) \\
Object 3a $\natural$   & 1.1 & -11.9 (18.7) &  16.5 (18.4) & 16.0 
 $\pm$ \tbn{L}13.9 & 43.2 & 70.2 & 145.9 & \cduo{$\pm$23.7} & 89 \\
Object 3a $\flat$   & 1.1 & -4.5 (20.5) &  20.5 (19.8) & 9.75 
 $\pm$ \tbn{L}21.4 & 45.5 & 67.1 & 134.2 & \cduo{$\pm$36.4} & 93 \\
\hline
MRC 0156$-$252 & 2.3 & -2.54 (7.02) &  1.09 (7.36) & 0.14 $\pm$ \tbn{L}4.28 &
 10.44 & 14.2 & 161.4 & \cduo{$\pm$59.4} & 13 \\
MRC 0156$-$252 & 2.7 & -0.53 (7.57) & -0.54 (8.33) & 0.04 \tbn{L,U} &
 4.19 & 0.89 & 60.1 & \cduo{$\pm$60.0} & (13) \\
MRC 0156$-$252 & 3.4 &  0.43 (8.44) & 0.81 (9.38) & 0.05 \tbn{L,U} &
 5.49 & 1.00 & 114.0 & \cduo{$\pm$60.0} & (16)\\
\hline
3C 441 a  & 3.1 & 3.89 (5.07) &-0.70 (4.97) & 1.03 $\pm$ \tbn{L}6.30 & 
	10.02 & 45.7 &  77.9 & \cduo{$\pm$50.5} & 22 \\
3C 441 B$\ddagger$ & 3.1 & 0.53 (4.66)  &-1.74 (4.69)   & 0.09 $\pm$ 
\tbn{L}2.80 & 6.90  & 14.0 &  46.5 & \cduo{$\pm$59.4} & 9 \\
3C 441 c$\ddagger$ & 3.1 & 1.26 (12.04) & 10.01 (11.42) & 3.53 $\pm$ 
\tbn{L}15.61 & 23.94 & 54.1 & 124.4 & \cduo{$\pm$45.1} & 57\\
3C 441 E  & 2.6 & 5.44 (12.28) & 19.02 (12.94) & 17.65 $^{+9.31}_{-8.08}$ &
        36.24 & 90.5 & 120.0 & \cduo{$\pm$14.9} & 81 \\
3C 441 F & 2.0 & 5.85 (19.20) & 16.41 (21.69) & 5.54 $\pm$ \tbn{L}27.54 &
        43.25 & 48.4 & 118.2 & \cduo{$\pm$47.6} & 99 \\
3C 441 G  & 2.6 &-0.05 (12.74) &-5.90 (13.13)  & 0.29 $\pm$ \tbn{L}9.90 &
        20.52 & 18.3 &  37.7 & \cduo{$\pm$59.2} & 31 \\
3C 441 H  & 2.3 &-12.06 (14.02)&-8.85 (14.83)  & 6.65 $\pm$ \tbn{L}22.22 &
        32.63 & 66.5 & 11.1 & \cduo{$\pm$37.5} & 87 \\ 
3C 441 S & 2.3 & 0.37 (1.97) &-0.23 (1.96) & 0.02 \tbn{L,U} &
        2.27  &  4.7 &  66.9 & \cduo{$\pm$59.8} & (7) \\
\hline
\end{tabular}

\medskip

Key: Source: Source name and component ($\natural$: natural image; 
$\flat$: `despiked' image; $\ddagger$: data is based on 
1995 observations only; S is a bright star below 3C 441 accessible on the
1997 frames only); r: radius of
photometry aperture (arcseconds);
 $q \pm \sigma_q, u \pm \sigma_u$: normalized Stokes
Parameters (per cent) with respect to 83\degr\ E of N;
 $P\pm\sigma_P$: percentage polarization (debiased) with 1$\sigma$ error
(\tbn{L} --- the $1\sigma$ lower limit for polarization is zero; \tbn{L,U}
--- the $1\sigma$ `confidence interval' is identically zero even though
the best point estimate polarization is non-zero);
$2\sigma$.UL: $2\sigma$ upper limit
(in per cent) for polarization in objects unlikely to be polarised;
Prob: the probability
that there is underlying polarization;
$\theta\pm\sigma_\theta$: Electric vector orientation E of N (\degr);
$\PHI{K}$: $1\sigma$ upper limit (per cent) to fraction of light arising
in the nucleus, or ($2\sigma$) limit where the $1\sigma$ limit is zero.
\label{restab} \end{minipage}

\end{table*}

 As is recommended for polarimetry, the normalised Stokes parameters $q$ 
and $u$ are tabulated; the reference axis for IRPOL2 is that $q>0, u=0$ 
corresponds to a polarization orientation of 83\degr\ E of N and that for 
$q=0, u>0$, the polarization orientation is 128\degr. The reference axis 
was verified by observing the standard star HD 215806 \cite{Whittet+92a}.

 In order to produce the pictures of the three sources presented here 
(Figures \ref{Wpic}, \ref{Mpic} and \ref{Cpic}) the upper and lower
images from the final mosaics from all the waveplate positions 
were combined to form unpolarised images with the maximum depth; earlier 
data were also included, where available, as noted in the figure captions. 

\section{Results and Discussion}

\subsection{The fraction of light arising in the nucleus}

 As discussed in L\&E, the fraction $\PHI{K}$ of the \Kb-band observed
light arising in the active nucleus is given by $\PHI{K} =
\Ptt{K}/\Pq{K}$, where $\Ptt{K}$ is the observed polarization and 
$\Pq{K}$ is the undiluted polarisation of the nuclear component scattered 
towards Earth. Models
\cite{Manzini+96a} suggest that the intrinsic polarisation for realistic
scattering by dust (at 90\degr) will be such that $1/\Pq{K} = 2.5 \pm 0.5$, 
while electron scattering at any angle greater than 45\degr\ will give 
$1/\Pq{W} = 2 \pm 1$. It follows that $\PHI{K} \la 3\Ptt{K}$, so we 
can calculate a $1\sigma$ upper limit on the fraction of \Kb-band light 
arising in the active nucleus as 3 times the $1\sigma$ upper limit 
$P+\sigma_{P}$ on 
the polarization. These values are also given in Table \ref{restab}.

\subsection{LBDS 53W091}

The $K$-band image of 53W091 which results from stacking all of our
data is shown in
Figure \ref{Wpic}. The companion object to the south-east of 53W091 appears
to be at the same redshift, and is labelled `$3a$' in accordance with
the labelling of Spinrad et al. \shortcite{Spinrad+97a}. The position of 
their third component 
is also marked (labelled `4'), although there is not a distinct source 
on our image.

Two stars on our image (the star on the top right of Figure \ref{Wpic} and a 
brighter one in the lower slot of the focal plane mask, not shown) were 
identified with stars whose B1950 co-ordinates were obtained from the {\em 
Digitized Sky Survey} \cite{Lasker+90a}. Offsetting from these stars, the 
B1950 co-ordinates of the \Kb-band sources were obtained and are given in 
Table \ref{positab}.

\begin{table}
\caption{B1950 co-ordinates of \Kb-band sources in 53W091 field}
\label{positab}
\begin{tabular}{lcc}
\hline
Source &  $\alpha$ & $\delta$\\
53W091 &  17h 21m 17\fs898 $\pm$ 0\fs057 
& $+50\degr 08\arcmin 48\farcs34 \pm 0\farcs29$ \\
3a & 17h 21m 18\fs156 $\pm$ 0\fs029 &
 $+50\degr 08\arcmin 46\farcs34 \pm 0\farcs57$\\
 
\hline   
\end{tabular}
 
\medskip

Key: $\alpha$: B1950 Right Ascension; $\delta$: B1950 Declination.
\end{table}

Spinrad et al. \shortcite{Spinrad+97a} discuss whether
$3a$ and 53W091 together form a system displaying the alignment effect, 
given that
the axis connecting the two objects is at a position angle of 126\degr, 
(comparable to the radio axis at 131\degr), and that
the diagonal distance between 53W091 and $3a$ is
4\arcsec, comparable to radio-lobe separation in the 4.86 GHz map
of Spinrad et al.  \shortcite{Spinrad+97a}.
However, the visible position of 53W091 (with which our infrared position
measurement is consistent) is midway 
between the two radio lobes (and thus presumably coincides with the radio
core), while the visible/infrared position of object $3a$ lies just 
outside the SE lobe. Thus, while an interaction between the (brighter) 
SE lobe and object $3a$
may be influencing the optical/radio appearance of the complex, 53W091
would appear to be the host galaxy of the central radio source.
Furthermore, the visible/ultraviolet 
spectra of both sources suggest they are dominated by old stars, implying
that both objects are distinct galaxies, and that any aligned light
associated directly or indirectly with the active nucleus of 53W091 makes
at most a minor contribution to the visible and infrared morphology of the 
complex. [This interpretation is supported by recent WFPC2/NICMOS imaging
of 53W091 \cite{Dunlop-99a}.]

Polarimetry was performed on both 53W091 and on Object $3a$; results for 
both are given in Table \ref{restab}. Since the objects were very faint and 
close together, the photometry aperture was not chosen according to the 
method in L\&E, but was set to a radius of 4 pixels ($1\farcs1$). We also 
attempted to prune the frames with the greatest noise from our 
data and repeated the polarimetric analysis. Results for both the natural 
($\natural$) and `despiked' ($\flat$) data are given in Table \ref{restab}.

 There is a weak indication that Object $3a$ is polarised, with a 70 per
cent chance of the polarisation being genuine. If there truly is 
polarisation at a level of 10-15 per cent, then 30-45 per cent of the 
light from object $3a$ could be scattered, and the source could consist 
entirely of scattered light within the error bars. (Dust scattering and 
non-perpendicular electron scattering will not result in total linear 
polarization of the scattered light.)

 Is it possible that a beam from 53W091 is being scattered by a cloud at 
the position of $3a$? The geometry suggests that this cannot be the case, since
the polarization orientation is around 140\degr, which is
nearly parallel to the line connecting $3a$ to 53W091. If $3a$ were a
knot induced by a beam emerging from 53W091, a polarization orientation 
nearer 30\degr\ would have been expected. 

The core of 53W091 itself provides no evidence for polarization, and it
would be very difficult to reconcile our data with 
a polarization as high as the 
40 per cent which Chambers (private communication) has suggested.
We can only conclude that this previous measurement was in error, and/or
was not based on sufficient integration time to properly constrain the
polarization of this source.  
Our measured polarization 
orientation is 38\degr, nearly perpendicular to the 
radio axis and line to object $3a$, 
but this is only marginally significant with errors of $\pm60\degr$ on our 
formal measure of the polarization angle.

\subsection{MRC 0156-252}

 The criterion used by L\&E to select the aperture for polarimetry did 
not yield a unique result for this object, so photometry is given in Table
\ref{restab} for apertures of radius 8, 10 and 12 pixels. In all cases 
the best point estimate of the polarization is less than 0.15 per cent; 
and for the 10 and 12 pixel apertures, the formal $1\sigma$ confidence 
interval indicates that the source is totally unpolarised. At most, if we 
assume that the polarization is due to the scatterering of light 
originally travelling perpendicular to our line of 
sight, 16 per cent of its \Kb-band light is due to this component, and 
might be assumed to be arising in an active nucleus.

The most recent radio maps of MRC 0156-252 \cite[4710 MHz and 
8210 MHz]{Carilli+97a} do not
display the marked head-tail asymmetry characteristic of beamed quasars.
Nevertheless, it is possible that MRC 0156$-$252 is an obscured quasar
\cite{Eales+96a}; if this is the case, we must be looking close to
`straight down the jet' with a shallower scattering angle for infrared
light. If this were the case, our scattering assumptions would be invalid 
and hence our upper limit for $\PHI{K}$ would be an underestimate. 

\subsection{3C 441}
 
 Our 1995 polarimetry (L\&E) of 3C 441 was inconclusive, so we took
advantage of this observing run to obtain further data.  Given the
uncertainty posited by Neeser \shortcite{Neeser-96a} over the
identification of 3C 441, and the interest in object $c$ of Lacy et al.
\shortcite{Lacy+98a}, we measured the polarization of all the objects on
the field. The 1995 and 1997 images were stacked together to obtain
polarization measurements of objects $a$ (the putative core of 3C 441) and
$E$ thru $H$; the 1995 data alone was used to obtain polarization data on
$B$ and $c$. 

 The only object with a strong indication (90 per cent chance genuine) of
polarization is $E$. The orientation is 120\degr, which would be roughly
parallel with the radio jet -- but the position angle which $E$ makes with
the presumed core $a$ is close to 0\degr, which means that a model of $E$
scattering light from $a$ is possible. It would not be neccessary for
light from $a$ to be beamed into $E$; if $E$ subtends only a small solid
angle of the light emitted by $a$, any light from $a$ scattered by $E$
would be quasi-unidirectional. 

\begin{table}
\caption{Polarization in 3C 441 in two epochs}
\label{vari441}
\begin{tabular}{lcc}
\hline
Epoch &  $p (\%)$ & $\theta (\degr)$\\
August 1995 &   $ 0.26 + 8 $ & $-5  \pm 59$\\
August 1997 &   $ 19 \pm 13 $ & $13  \pm 19$\\
\hline   
\end{tabular}
 
\medskip

Key: $p$: Debiased polarization; $\theta$: polarization orientation. 
\end{table}

The availability of data from two epochs makes a crude test of temporal
variability possible. The data are presented in Table \ref{vari441}, and
it is clear that the measurements in the two epochs are consistent within
the errors. Variation of polarization over a two year span can neither be
proven nor disproven. If it is safe to pool the data from both epochs, the
best estimate polarization, for the presumed radio galaxy at $a$, is 1 per
cent at 78\degr;  but there is a 54 per cent chance that $a$ is
unpolarised with this result being merely an artefact of the noise. Even
our result for $E$ has a ten percent chance of being a noise-induced
spurious result. 

\subsection{Polarized companions?}

 The presence of the 3C 441 case -- i.e.\ where a companion object seems 
to be
polarized in a manner consistent with the scattering of light emanating
from the radio-loud object -- may be significant in the wake of recent
high-resolution imaging polarimetry of high redshift radio galaxies. Tran
et al.\ \shortcite{Tran+98a} used the Keck I to obtain extended imaging
polarimetry of 3C 265, 3C 277.2 and 3C 324. In all three cases, the
polarization maps displayed bipolar fans of polarization vectors centred
on the nucleus, perpendicular to the optical structure and misaligned by
tens of degrees with the radio axis. Earlier structural information on one
radio galaxy was obtained by di Serego Alighieri, Cimatti \& Fosbury
\shortcite{Alighieri+93a}; their $V$-band polarimetry of the
$z=0.567$ object $1336+020$ showed perpendicular polarization in
a northern knot and in extended emission, higher than in the core. 

Contour maps of the three Tran et al.\ \shortcite{Tran+98a} sources are
provided, at levels relative to the peak intensity of the the central
knot, and all three sources include companion objects. Little or no
polarization is seen in the bright ($\sim 20$ per cent of peak) companions 
of 3C
277.2 and 3C 324. But in 3C 265, a faint companion object also exhibits
the polarization seen in the fan -- the object is a knot less than 8 per 
cent of peak intensity and lies beyond the extension of the $V$-band optical
structure, in the same direction but unconnected with the optical core in
contours down to 2 per cent of peak. Such a faint polarized knot could 
readily be
identified with light redirected by a cloud of scattering particles. 

 Assuming an $\Omega_0 = 1.0, \Lambda=0$ cosmology with $H_0 = h_0$ 
km\,s$^{-1}$\,Mpc$^{-1}$, $h_0 = 100$, the knot in 3C 265 which lies about 
9\arcsec\ from 
the core, is separted from the core by about 36$h_0$ kpc. The extended 
stucture of $1336+020$ of about 3\arcsec\ corresponds to 12$h_0$ kpc. In 
comparison the 53W091 to $3a$ separation and the distance 
between 3C 441 $a$ and $E$ both correspond to approximately 16$h_0$ kpc. 
[Angular to linear scale conversion factors are taken from Peterson 
\shortcite[Fig. 9.3]{Peterson-97a}.] So the structure of these companion 
objects is of comparable scale to those in the literature.

 It is possible, therefore, that companion $E$ to 3C 441 is an 
illuminated object scattering light in the manner of the extended 
structure seen in 3C 265 and $1336+020$. A determination of the redshift 
of object $E$ would be required, however, to confirm that its presence in 
the field is not merely a line-of-sight effect. The parallel polarization 
of companion $3a$ to 53W091 cannot be explained in this way; but the 
measurement is too marginal to invite a search for parallel 
polarization mechanisms without more examples being found first.

\section{Conclusions}
 
 Polarimetry of faint objects requires long integration times. The time 
available has permitted us to rule out the existence of very high 
polarizations in all the objects studied, at least for light emitted 
along the line of 
sight to Earth. It would obviously still be possible for light emitted in other 
directions from these objects to be polarised. `Polarization' mentioned 
in the following conclusions should be understood in the restricted sense 
of light leaving the source in the direction of Earth. Under the Unified 
Model, radio galaxies (a class of AGN assumed to be oriented with their 
jets perpendicular to that line of sight) would be more likely to display 
polarization originating in scattering or synchrotron radiation in the 
light travelling Earthwards than in directions closer to the jet.

 In the case of MRC 0156$-$252, 
which lies beyond a virtually dust-free part of our own Galaxy, we can be 
reasonably certain that this radio galaxy is not polarised, and the 
\Kb-band light has not been scattered before reaching us. If some of 
the \Kb-band light has originated in the active nucleus, its 
contribution should be smaller than at visible wavelengths 
\cite{Manzini+96a}; this being the case, 
subtraction of our image or a synthetic symmetrical galaxy could well 
reveal the structure of the active component at visible wavelengths, 
given the visible structure observed by McCarthy et al. 
\shortcite{McCarthy+92a}.

 In LBDS 53W091, we can rule out the contribution of an active nucleus to 
providing more than $\sim 25$ per cent of the observed light. The 
majority of its $K$-band light, therefore, must be 
presumed to be due to its stellar population, and its $R-K$ colour
remains consistent with an age in the range $2.5-3$ Gyr.
The nature of its companion object $3a$, possibly polarised and of 
unclear physical relationship with 53W091, warrants further investigation.

 In 3C 441, the polarization from object $E$ may indicate that $E$ is 
scattering light from $a$ (whose identification as the central engine would 
thus be vindicated); the orientation of $E$'s polarization would not be 
consistent with the source being located within $E$ or $F$ and emitting jets at 
$\sim 145\degr$. Therefore, we favour the traditional identification of 
the central engine with $a$.

 The presence of companion objects perpendicularly polarized to the
line of sight may be providing our first hints of
a near infrared scattering medium concentrated in clouds in the
vicinity of active nuclei. Further studies are required, however, to
establish the nature of particular companions, and the presence of near
infrared companions as a whole.

\subsection*{Acknowledgements}

 We thank Antonio Chrysostomou for help with the observations, and an
anonymous referee for useful suggestions. 
The United Kingdom Infrared Telescope is operated by the Joint Astronomy 
Centre on behalf of the U.\,K. Particle Physics and Astronomy Research 
Council. We thank the Department of Physical Sciences, University 
of Hertfordshire for providing IRPOL2 for the UKIRT.

 This research has made use of the {\sc nasa/ipac} extragalactic database
({\sc ned}) which is operated by the Jet Propulsion Laboratory, CalTech, 
under contract with the National Aeronautics and Space Administration. 
Data reduction was performed with {\sc starlink} and {\sc iraf} routines. We
acknowledge the use of NASA's {\em SkyView} facility
({\tt http://skyview.gsfc.nasa.gov}) located at NASA Goddard
Space Flight Center. GL thanks {\sc 
pparc} for a research award.

\end{document}